\documentclass[manuscript,acmtog]{acmart}
\AtBeginDocument{%
  }

\acmDOI{}

\usepackage{color}
\usepackage{soul}    
\usepackage[capitalize,nameinlink]{cleveref}

\graphicspath{ {figures/} }
 
\usepackage{gensymb} 

\usepackage{enumitem} 

\usepackage{xspace} 

\begin{document}

\title{Symbiotic AI: Augmenting Human Cognition from PCs to Cars}

\author{Riccardo Bovo}
\affiliation{%
  \institution{Imperial College London}
  \city{London}
  \country{UK}
}
\email{rbovo@google.com}

\author{Karan Ahuja}
\affiliation{%
  \institution{Northwestern University}
  \city{Evanston}
  \state{Illinois}
  \country{USA}
}
\email{kahuja@google.com}

\author{Ryo Suzuki}
\affiliation{%
  \institution{University of Colorado Boulder}
  \city{Boulder}
  \state{Colorado}
  \country{USA}
}
\email{ryo.suzuki@colorado.edu}

\author{Mustafa Doga Dogan}
\affiliation{%
  \institution{Adobe Research}
  \city{Basel}
  \country{Switzerland}
}
\email{doga@dogadogan.com}

\author{Mar Gonzalez-Franco}
\affiliation{%
  \institution{Google}
  \city{Mountain View}
  \state{}
  \country{USA}
}
\email{mgonzalezfranco@google.com}

\renewcommand{\shortauthors}{Bovo et al.}

\begin{abstract}
As AI takes on increasingly complex roles in human-computer interaction, fundamental questions arise: how can HCI help maintain the user as the primary agent while augment human cognition and intelligence? This paper suggests questions to guide researchers in considering the implications for agency, autonomy, the augmentation of human intellect, and the future of human-AI synergies. We observe a key paradigm shift behind the transformation of HCI, shifting from explicit command-and-control models to systems where users define high-level goals directly. This shift will be facilitated by XR technologies, whose multi-modal inputs and outputs offer a more seamless way to convey these goals. This paper considers this transformation through the lens of two cultural milestones: the personal computer and the automobile, moving beyond traditional interfaces like keyboards or steering wheels and thinking of them as vessels for everyday XR.  
\end{abstract}




\maketitle

\section{Introduction}

Artificial Intelligence (AI) has rapidly evolved from specialized applications to an omnipresent technology, with Large Language Models (LLMs) reshaping how we work, communicate, and move. AI is not just able to recognize complex patterns and interpret unstructured data, but also generate sophisticated outputs to match human expectations (to a degree or another)~\cite{zha2023data}. This generative capability fuels a fundamental move away from rigid, command-and-control interfaces towards goal-oriented interactions. Soon, users will be able to simply convey their intentions, and an AI agent will autonomously determine the necessary steps to achieve them. This approach has already been used to break down complex scientific tasks into smaller, more manageable modules~\cite{zenil2023future}. The shift towards expressing intent rather than dictating specific commands may foster deeper cognitive argumentation, as users are challenged to articulate their goals clearly and critically evaluate the AI's proposed solutions~\cite{yakura2024empirical}. However, while these new LLMs are faster at solving complex tasks than "all" humans \cite{jumper2021highly}, and can outperform the “average” human in many tasks, they often do not exceed the abilities of "top" experts~\cite{brodeur2024superhuman}. LLM`s introduces fundamental challenges for the Human-Computer Interaction (HCI) community at large when interacting with AI. First, \textit{how can we ensure that users \textbf{remain the principal agents}}, retaining meaningful control even as AI handles intricate tasks? Second, \textit{how can we design AI interaction systems that \textbf{expand human cognition} and \textbf{intelligence}}?  This includes exploring how AI can effectively share expertise to help users bridge the gap between “average” performance and near-expert or expert-level capabilities—and simultaneously elevate their own cognitive and communicative skills. Realizing such a vision would require dynamic dialogue and iterative exploration, i.e., \textit{true} human-AI collaboration rather than one-shot question–answer formats. Addressing these questions will critically shape the future of human-AI collaboration, where, as Marshall McLuhan famously observed, 'we shape our tools, and thereafter our tools shape us'.To ground our discussion, beyond XR, we turn to two cultural milestones: \textit{personal computers} (PCs) and \textit{automobiles}. PCs exemplify tools of knowledge work, while cars provide an interface to the physical world, shaping how we navigate and experience our environment. Furthermore, cars have already become a sort of agent when combined with GPS, where we delegated decision making. We scape our minds on a traffic jam. We need to regain control later. We can see diametric learnings from these two technologies that are relevant for XRAI. Indeed AI is already transforming both domains, from large language models (LLMs) serving as "co-pilots" for writing~\cite{sirisathitkul2024slow}, coding~\cite{arjun2024beyond}, and content creation, to advanced agents or driver-assistance systems that augment the driving experience~\cite{rana2024automotive}.  
By considering the implications for agency, autonomy, and cognitive augmentation, we can help shape the development of XRAI systems that empower humans while respecting their fundamental need for control and understanding.
We propose that HCI, particularly with XR technologies, lies at the heart of answering these questions. By blending multi-modal inputs (e.g., speech, gaze, gestures) with immersive outputs (e.g., augmented dashboards, stand-up displays, near-field-of-view screens), XR makes it more \textit{intuitive} to convey high-level goals and to interpret AI-driven feedback in real time . From simplifying how we express our intentions to offering vivid, context-aware information overlays, XR has the potential to bridge the gap between the nature of human cognition and the nature of AI~\cite{Bovo_2025_emBARDiment}.
XR also makes it possible to visually convey to the human the sparks of shared "theory of mind" with the AI: what is the AI thinking or how did it arrive to that conclusion.

\section{The End of Command-and-Control Model}

A shift to abstract goals carried out by agents. If you have a goal (like sending an email), you need to decompose that goal into the set of interactions required to achieve it (open \textit{Outlook}/\textit{Gmail}, write an email, enter an address, send the email). In the command-and-control model, users give explicit, specific commands, and the computer executes them exactly as instructed. This interaction style establishes a clear hierarchy where the user has full control over the computer's actions. A fundamental pattern we have seen with generative AI is interfaces that are iterative and allow users to adjust their aims in subsequent iterations. This represents a shift away from the deterministic and rigid command-and-control model to one where interactions are more conversational, iterative, and adaptive.

\begin{enumerate}[leftmargin=0.5cm]
    \item \textit{How and when do users retain control while the agents perform steps toward the goal without losing the advantages of having an agent perform those steps?}
    \item \textit{How many agents can one supervise at the same time? What is the ideal number of agents? What are the tasks they help accomplish?}
    \item \textit{What is the trade-off between direct control and agent supervision?}
    \item \textit{What are the agent representations, and how do we represent their work?}
\end{enumerate}

We argue it might be more related to the task the user is trying to accomplish, the context, and the purpose: productivity, gaming, social. However, these questions have already arise in the space of self-driving cars and learnings could be drawn from there ~\cite{huang2023v2x}. 

\section{The End of Computerization}
A primary need for PCs is computerization, defined as ``the computer needs the information processed in a form that a computer can use'' \cite{kling1996computerization}. If you have some data, you need to enter it into a spreadsheet using your keyboard (i.e., you computerize that data). For example, manually counting and recording inventory items in a spreadsheet for a retail store remains a tedious and manual computerization task, often requiring physical verification and data entry. AI leads to a paradigm where data does not need to be ``computerized'' in the traditional sense because AI can understand and process data in its natural, unstructured form. For AI PC systems that process unstructured data in real-time, edge computing capabilities may be essential, or hardware that can efficiently process unstructured data locally (LPU), reducing the need for constant data transmission.

\begin{enumerate}[leftmargin=0.5cm]
    \item  \textit{How and when do we allow unstructured data to flow into AI, and how do we make such a flow as intuitive and easy to control as possible?}
    \item \textit{How will the hardware of the AI PC need to incorporate advanced security features to protect sensitive unstructured data?}
    \item  \textit{What is the split between local and edge computing capabilities?}
    \item \textit{What are the values of existing UI primitives? How can UI primitives be AI-embedded?}
\end{enumerate}

\section{The End of Humans Behind the Wheel}

Cars are machines that are pervasive in our lives—arguably even more than laptops, and certainly more than XR. More importantly, cars serve as an interface between us and the rest of the world, connecting our actions to the physical and social environments we navigate daily. For over a century, the car has required humans to sit in the driver’s seat, physically operating the vehicle. However, this traditional relationship between humans and cars is now evolving ~\cite{rana2024automotive}. While AI and automation promise fully autonomous driving, the future may instead focus on augmenting the driver’s capabilities rather than replacing them entirely. In this new paradigm, humans are no longer mere drivers —they become collaborators with the car ~\cite{huang2023v2x}. AI copilots and advanced sensory systems provide assistance, adapt to context, and enhance the driving experience without fully removing human agency. The result is a shift where humans remain “in control” but are supported by a symbiotic relationship with AI. This transition redefines what it means to “drive,” incorporating shared attention systems, Multimodal interfaces, and cognitive augmentation, creating an environment where humans can still be in the driver’s seat—just not as we’ve historically imagined it.

\begin{enumerate}[leftmargin=0.5cm]
    \item  \textit{Indoor car perception meets outdoor car perception}. To achieve a real symbiosis of Humans and cars, it is clear that we need to achieve more human behavior understanding, highlighting the role of tags and sensors fusion. To achieve shared attention co-piloting with salient AI.

    \item \textit{Connective experiences}. Fleet socialization and the “Hive effect.” What does a true connected car mean? What is the “social media of a car”? Which forces act on me, and how do I act on others (e.g., congestion pricing, crowdsourcing parking spots, carbon footprint, road rage, battery charge)?
    
    \item \textit{Large Language Models (LLM) and Computer Vision (CV)}. Advanced driver features such as Geo-spatial anchoring of items, spatial AR windshield, or a smart dashboard will be enabled via AI multimodal I/O in cars combined with natural interactions and full stack applications and SDKs.
    
\end{enumerate}

\section{Conclusion}
AI's transformative impact, evident in PCs and cars, demands a redefined HCI, especially within XR's immersive landscape. Crucially, we must address agency, control, and cognitive augmentation to ensure XR empowers users, not eclipses them. This requires intuitive, multimodal interfaces and transparent AI, built through collaborative efforts between research and industry. Ultimately, success lies not just in enhancing capabilities, but in fostering user understanding and control within an AI-infused world.





\bibliographystyle{ACM-Reference-Format}
\bibliography{sample}


\end{document}